\documentclass{article}

\usepackage{arxiv}

\usepackage[utf8]{inputenc} 
\usepackage[T1]{fontenc}    
\usepackage{hyperref}       
\usepackage{url}            
\usepackage{booktabs}       
\usepackage{amsfonts}       
\usepackage{nicefrac}       
\usepackage{microtype}      
\usepackage{lipsum}		
\usepackage{graphicx}
\usepackage{natbib}
\usepackage{doi}
\usepackage{amsmath}

\title{Generating tailored high frequency features in core collapse supernova gravitational wave signals applicable in LIGO interferometric studies}

\author{ \href{https://orcid.org/0000-0000-0000-0000}{\includegraphics[scale=0.06]{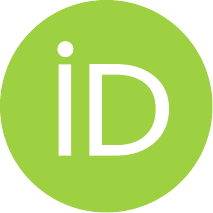}\hspace{1mm}César Tiznado}\thanks{Claudia Moreno - Correspoding author, claudia.mgonzalez@academicos.udg.mx.} \\
	Escuela de Ingeniería y Ciencias\\
	Tecnologico de Monterrey\\
        Monterrey, N.L., 64849, México \\
	\texttt{A00838226@tec.mx} \\
	\And
	\href{https://orcid.org/0000-0000-0000-0000}{\includegraphics[scale=0.06]{orcid.pdf}\hspace{1mm}Alejandro Casallas-Lagos} \\
	Escuela de Ingeniería y Ciencias\\
	Tecnologico de Monterrey\\
        Monterrey, N.L., 64849, México \\
	\texttt{alejandro.casallas@tec.mx} \\
        \And
	\href{https://orcid.org/0000-0000-0000-0000}{\includegraphics[scale=0.06]{orcid.pdf}\hspace{1mm}Javier M. Antelis} \\
	Escuela de Ingeniería y Ciencias\\
	Tecnologico de Monterrey\\
        Monterrey, N.L., 64849, México \\
	\texttt{mauricio.antelis@tec.mx } \\
        \And
	\href{https://orcid.org/0000-0000-0000-0000}{\includegraphics[scale=0.06]{orcid.pdf}\hspace{1mm}Claudia Moreno} \\
	Departamento de Física\\
	CUCEI, Universidad de Guadalajara\\
        Guadalajara, Jal., M\'exico \\
	\texttt{claudia.mgonzalez@academicos.udg.mx} \\
}

\renewcommand{\shorttitle}{\textit{GW from CCSN signals} }

\hypersetup{
pdftitle={Generating tailored high frequency features in core collapse supernova gravitational wave signals applicable in LIGO interferometric studies},
pdfsubject={q-bio.NC, q-bio.QM},
pdfauthor={Alejandro Casallas-Lagos},
pdfkeywords={Gravitational waves},
}

\begin{document}
\maketitle

\begin{abstract}
In this article, we introduce a methodology based on an analytical model of a damped harmonic oscillator subject to random forcing to generate transient gravitational wave signals. Such a model incorporates a simulated linear high-frequency component that mirrors the growing characteristic frequency over time observed in numerical simulations of core-collapse supernova gravitational wave signals. Unlike traditional numerical simulations, the method proposed in this study requires minimal computational resources, which makes it particularly advantageous for tasks such as data analysis, detection, and reconstruction of gravitational wave transients. To verify the physical accuracy of the generated signals, they are compared against the amplitude spectral of current LIGO interferometers and a 3D numerical simulation of a core-collapse supernova gravitational wave signal from the Andresen et al. 2017 model s15.nr. The results indicate that this approach is effective in generating scalable signals that align with LIGO interferometric data, offering potential utility in various gravitational wave transient investigations.
\end{abstract}


\section{Introduction}
Since Einstein predicted the existence of gravitational waves (GW) in 1916 \cite{Einstein:1916cc}, their search was a challenge that lasted for a century. Throughout this period of time, extensive research efforts have been dedicated to their discovery.
In 2015, the first experimental confirmation of a GW signal, associated with a black hole compact binary coalescence (CBC) system, known as GW150914 \cite{LIGOScientific:2016emj}, was achieved through the efforts of the LIGO (L1, H1) ~\cite{TheLIGOScientific:2014jea} and VIRGO~\cite{TheVirgo:2014hva} collaborations; later,  KAGRA ~\cite{Aso:2013eba} joined the search for GW, and together they are called the LVK collaboration.
Today, the LVK Collaboration has successfully detected more than one hundred GW events, making this remarkable achievement the result of dedicated efforts in theoretical \cite{Powell_2019}, observational \cite{Aerts_2021}, computational \cite{Andresen_2017} and experimental research \cite{yakunin2015gravitational}. 
After the first detection of a GW from a CBC, new feasible sources of GW are expected to be detected, such as GW signals from core collapse supernovae (CCSN) \cite{Kuroda_2016, Powell_2020, moore2014gravitational}. The first detection of a GW from CCSN will become the next great achievement in the history of GW research. 
CCSN are complex stellar explosions produced in the final stages of life of massive stars, whose masses are beyond 8M$\odot$, characterized by a large emission of energy through photons \cite{Aerts_2021}, neutrinos \cite{Powell_2020}, and GW \cite{M_ller_2019}. 
They are exceptional systems that involve several physical phenomena in a single event, which occurs at an extremely low rate in our Galaxy (approximately two occurrences per century \cite{Handler_2013}. CCSN are more complex systems than CBC due to their inherent physical complexity, which involves theoretical and experimental aspects that remain unknown today, such as the physical interactions that lead to such explosions, in particular microphysics beyond the nuclear mass limit \cite{Kuroda_2018, M_ller_2012, M_ller_2013}.
Thanks to advanced numerical simulations \cite{M_ller_2017, Powell_2019}, working together with computational data \cite{murphy2009model}, and laser interferometric advances, it is currently accepted in the GW astronomy community that; the GW signal emitted by a CCSN is (1) essentially stochastic \cite{Kuroda_2016, cerda2013gravitational, Kuroda_2018, M_ller_2019}, (2) has a strain amplitude, $h(t)$, in the order of $10^{-21}$m to $10^{-23}$m \cite{M_ller_2012, M_ller_2020, murphy2009model}, and (3) exhibit distinct physical components, features, incorporated in their signal, which describe the interactions that take place in different regions of the source \cite{astone2018new, yakunin2015gravitational, morozova2018gravitational, moore2014gravitational}. These characteristics can be regarded as deterministic and can be calculated to acquire physical data from the source of GW radiation.  
\\ \\
In recent years, several computational codes have been implemented to classify, determine, and characterize the different features incorporated in a GW CCSN signal. Their characteristics are given by specific physical interactions within the GW source, including hydrodynamic instabilities in the convection zone \cite{morozova2018gravitational, cerda2013gravitational}, buoyancy forces caused by gravity \cite{Aerts_2021} and stellar overshooting \cite{Handler_2013}, among others. 
All these components and characteristics associated with CCSN GW signals could be classified as deterministic along their time-frequency evolution signal, as demonstrated in \cite{M_ller_2013, M_ller_2017, Kuroda_2018}. 
An ongoing challenge within the study of CCSN is the quantification of these characteristics based on their physical behavior and the establishment of connections with the source parameters. This endeavor aims to reconstruct the essential properties of the GW progenitor. 
Due to the high computational cost of numerical simulations of CCSN GW signals, we are motivated to implement a method that generates CCSN GW signals using an analytical method, whose numerical solution simulates a GW signal with a continuous, linear, and strictly increasing High Frequency Feature (HFF) evolution \cite{astone2018new}.
We develop a simple but powerful computational model with low computational cost, making use of a theoretical model based on a second-order, nonhomogeneous, differential equation with random forcing to simulate the physical properties contained in the GW CCSN signal with presence of HFF, and a computational pipeline, based on the theoretical model \cite{astone2018new}, to emulate the time-frequency evolution of a GW CCSN signal.
Once the model has been computationally adapted, we perform variations in the parameters that define the theoretical model, producing signals with time-frequency evolution spectrograms with different HFF. Finally, we compared the generated GW strain signals with a 3D CCSN GW \cite{Andresen_2017} signal to contrast our results with the sensitivity curves associated with the LIGO interferometric data. 
One of the potential applications of this study includes the use of data analysis methods, the estimation of the HFF, and the interaction of computational physics to gain insight into the general physical attributes of this feature, which introduces a classical mechanical system.
\\ \\
This paper is organized as follows. In Section 2 we describe the physics of CCSN. Section 3 we explain the model of generation of GW through the damped harmonic oscillator. Section 4 focuses on the parameters that control the HFF of the simulations and the numerical scheme to solve the differential equation. The results of the GW simulation for CCSN are described in Section 5. In Section 6 we discuss the detectability of our GW simulations in the LIGO range. Finally, in Section 7 we present our conclusions.

\section{The physics of the core collapse supernova high frequency feature} \label{section:HFF}
%
After the CCSN an extremely compact and dense astrophysical object is born, a Proto-Neutron-Star (PNS). These stars, modeled by the laws of hydrodynamics, can be analytically described as a continuous and compressible mixture of gas and radiation. Its physical properties are induced by different interactions such as rotation, mass, magnetic fields, and tidal forces \cite{Aerts_2021, Handler_2013, murphy2009model}. Over time, PNS show dynamic pulsations, which are seen as areas that contract and expand on their photospheres, the external layer from which light is emitted \cite{Aerts_2021}. These pulsations result from standing waves propagating inside the star that constructively interfere with themselves, generating resonant modes \cite{cerda2013gravitational}. In essence, a star resembles a musical instrument, playing a unique tune composed of various modes dictated by its physical composition. 
The ``music" emitted by these astrophysical objects helps us to estimate the properties of the GW sources by detecting periodic surface movements. It is widely accepted in the current state of the art of GW astronomy that the main source of CCSN GW signals is the excitation of PNS modes during post-bounce evolution, which displays prominent stochastic behavior \cite{Andresen_2017, Kuroda_2016, M_ller_2017, Powell_2019}.
The frequencies of PNS oscillations depend on several factors within the stellar structure, including density, temperature, and gas motion \cite{morozova2018gravitational, murphy2009model}, the amplitude is determined by the magnitude of the excitation and damping processes, which can involve convection-induced turbulence and magnetic fields.
\\ \\
Pressure-driven oscillations, known as p-modes, are essentially acoustic waves. These modes are characterized by their restoring mechanism, which arises from pressure gradients within the stellar medium; p-modes are commonly observed in stars such as the Sun and solar-like stars \cite{Aerts_2021}. They tend to exhibit relatively high frequencies, with solar p-modes, for instance, falling within the frequency range of 1000 to 5000 $\mu$Hz, corresponding to periods ranging from 17 to 3 minutes, \cite{Handler_2013}.
However, gravity-driven oscillations, known as g-modes, are associated with buoyancy as the primary restoring force. In this case, gravity acts as the dominant force, restoring motion through buoyancy forces acting on variations in density across horizontal surfaces within the star \cite{astone2018new, cerda2013gravitational}. The resulting g-modes manifest themselves as standing internal gravity waves. Notably, some of the most extensively studied examples of g-modes are found in white dwarf stars \cite{Handler_2013}. 
\\ \\
Three main CCSN features have been discussed in current GW astronomy state-of-the-art: the Standing acceleration shock instability (SASI), the HFF, also known in modal analysis as g-mode, and memory \cite{M_ller_2019}. Each feature is in correlation with a particular physical mechanism that occurs during the emission of GW, given a specific response in the time-frequency evolution \cite{M_ller_2020}, 
in particular: (1) the HFF is related to physical properties located in the star core, and it is present in all GW CCSN simulations \cite{Andresen_2017, Kuroda_2016, Kuroda_2018,M_ller_2012, M_ller_2013, morozova2018gravitational}. It can be identified in a spectrogram as a progressively rising parabolic pattern starting around $150$~Hz and extends up to $700$~Hz \cite{2015ApJS_219_24O}; (2) SASI is a feature related to neutrino emission, which can be recognized in a spectrogram as a horizontal pattern around $120$~Hz to $200$~Hz, starting right after the HFF ends \cite{Powell_2019}; (3) memory is a low-frequency feature related to neutrino emission, can be recognized in a spectrogram as a bump starting around $1$~Hz growing up to $10$~Hz \cite{Powell_2020}.

\section{Generating core collapse supernovae gravitational waves signals: The damped harmonic oscillator} \label{section: Generation of GW}
%
In this section, we introduce an analytical approach developed for producing CCSN GW signals \cite{cerda2013gravitational}. The time-frequency plots of these signals demonstrate a linear increase in frequency over time, unlike those from numerical simulations. This analytical method can generate time-frequency responses that closely match their numerical counterparts in the CCSN GW signal and can be adjusted to accommodate the LIGO interferometric noise level. This flexibility makes it suitable for various GW studies that involve data analysis, parameter estimation, detection, and reconstruction with a low computational cost compared to CCSN numerical simulations.  
\\ \\
We extract the physical characteristics of an oscillatory system described through a second-order, linear, nonhomogeneous differential equation that incorporates random forcing \cite{landau2013course}. Subsequently, we adapt these properties computationally to simulate the generation of GW signals.
The main property of these generated signals is the presence of the HFF in its time-frequency spectrogram that describes a clear linear rise in time.
As we expect, using this generated model, we can study the physical properties of the HFF, controlling its evolution at the first-order approximation.
$\omega$, and time duration $t$, among others. 
Once the model has been defined, we describe its numerical solution using the semi-implicit Euler method and all the properties induced from the solution.  
\\ \\
A second-order linear differential equation with constant coefficients \cite{landau2013course} has found widespread application in physics, particularly in the modeling, vibrating, oscillatory, and resonant systems. These systems encompass mechanical, wave-based and electrical phenomena, ranging from the fundamental simple harmonic oscillator to more intricate variants that incorporate damping and external influences introduced by driving forces. In its homogeneous form, the linear and second-order differential equation with constant coefficients, which describes an oscillating system, can be expressed as follows \cite{astone2018new}:
 \begin{equation}\label{equ:2ndOrderDiffEqu2}
    \frac{d^{2} h(t)}{dt^{2}} + \frac{\omega_0}{Q} \frac{d h(t)}{dt} + \omega_0^{2} h(t) = 0 \; ,
 \end{equation}
where $h(t)$ denotes the time dependent solution, being $t$ the time duration of the signal. $\omega_0$ denotes the natural, i.e., undamped, angular frequency of the system; and $Q$ the quality factor associated with the energy content of the system. It is widely recognized that the solution to such a differential equation serves to elucidate three distinct types of mechanical oscillations induced by damping in physics. These oscillations are commonly termed underdamped, overdamped, and critically damped, and their categorization is based on the characteristics of their respective polynomials.
Unlike the underdamped oscillations, in both overdamped and critically damped cases, the system evolves under the effect of a higher disruptive force, avoiding any oscillation in the process. In all three different scenarios, the equation has an analytical solution that can be found by implementing classical differential equation methods. Underdamped oscillations are distinguished by their gradual decrease in amplitude over time, a behavior influenced by the presence of a relatively lighter external force acting on the system. However, as previously mentioned, a GW CCSN signal is a stochastic process. Therefore, this characteristic must be taken into account within the analytical model.
To accomplish this, we will utilize an extended version of equation \ref{equ:2ndOrderDiffEqu2}, which incorporates an additional stochastic component known as \emph{Driving Force}, denoted $s(t)$. This inclusion is essential to reproduce the stochastic property of the GW CCSN signal.
The force is mathematically modeled by a Dirac delta function as follows:
\begin{equation}\label{E:Forcing}
     s(t) = s_n \delta(t-t_n),
\end{equation}
in which $t_n$ with $n=1, 2, \dots, N$ are $N$ times instants where the driving force is applied to the system, and $s_n$ is the random amplitude of the force which is in the interval $[s_{min},s_{max}]$. 

\begin{figure}
    \centering
    \includegraphics[scale=0.5]{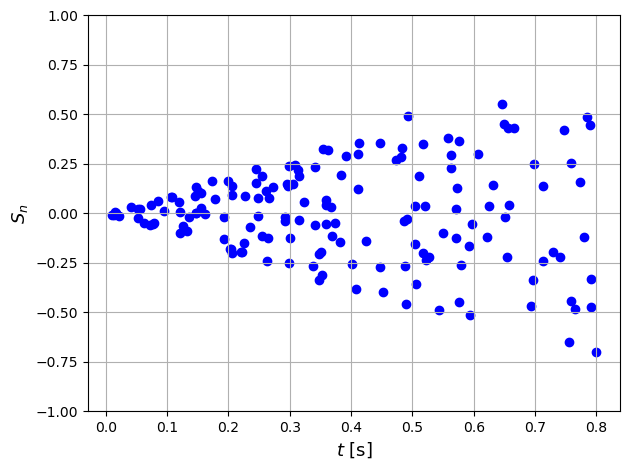} %
    \caption{\footnotesize Random forcing modeled in terms of the amplitude distribution $s_{n} = [-1,1]$ in a time interval $t = [0.0,0.8]$ with 100 elements. Each parameter dictates the amplitudes reached by the waveform during the time window and its distribution.
     }
    \label{fig:amp_an}
\end{figure}

The physical system that we will model from a mechanical point of view as a damped harmonic oscillator with a random force $s(t)$, is represented by the equation:
\begin{equation}\label{E:Driving}
        \frac{d^ 2 h(t)}{d t^{2}} + \frac{\omega(t)}{Q}\frac{d h(t)}{d t} + \omega(t)^{2}h(t) = s(t), 
\end{equation}
where the force is responsible for driving the evolution of the system as an external agent that controls the dynamical evolution. For example, its solution can take the form of a continuous periodic function of time, which is a common occurrence in resonance frequency problems. This theoretical approach makes use of fundamental building blocks to model the nature of the signals,
enclosing: (1) the quality factor $Q$, which plays a pivotal role in regulating the energy stored within the system and its subsequent dissipation over time due to damping. This factor will be employed to manage the linear growth observed in the spectrogram of the GW signal;
(2) the driving force, which model the stochastic component present in this radiation. By modifying the parameters included in the differential equation, we can create generated spectrograms that mimic the GW signals, each characterized by distinct linear evolutions of the HFF.
The characteristics that were previously established come together in a manner that functions as a mechanical representation of a PNS that expands and contracts in a repetitive manner, resembling the oscillations of the mechanical system and dissipating energy with each cycle. The damping component symbolizes the resistance against the surface as it attempts to alter its volume during each expansion and contraction cycle \cite{morozova2018gravitational}.
The forcing is modeled with an amplitude interval $s_n$ that is uniformly distributed, including the Dirac delta function positioned at values of $t_{n}$, see Fig. \ref{fig:amp_an}. These values $t_{n}$ correspond to instantaneous accelerations, driven by a frequency associated with the rate of triggers per unit of time.
%
\section{Parameters that controls the generated core collapse supernova high frequency feature over time} \label{subsection: Phen gmode -model}
\begin{figure}
     \centering
     \includegraphics[scale=0.4]{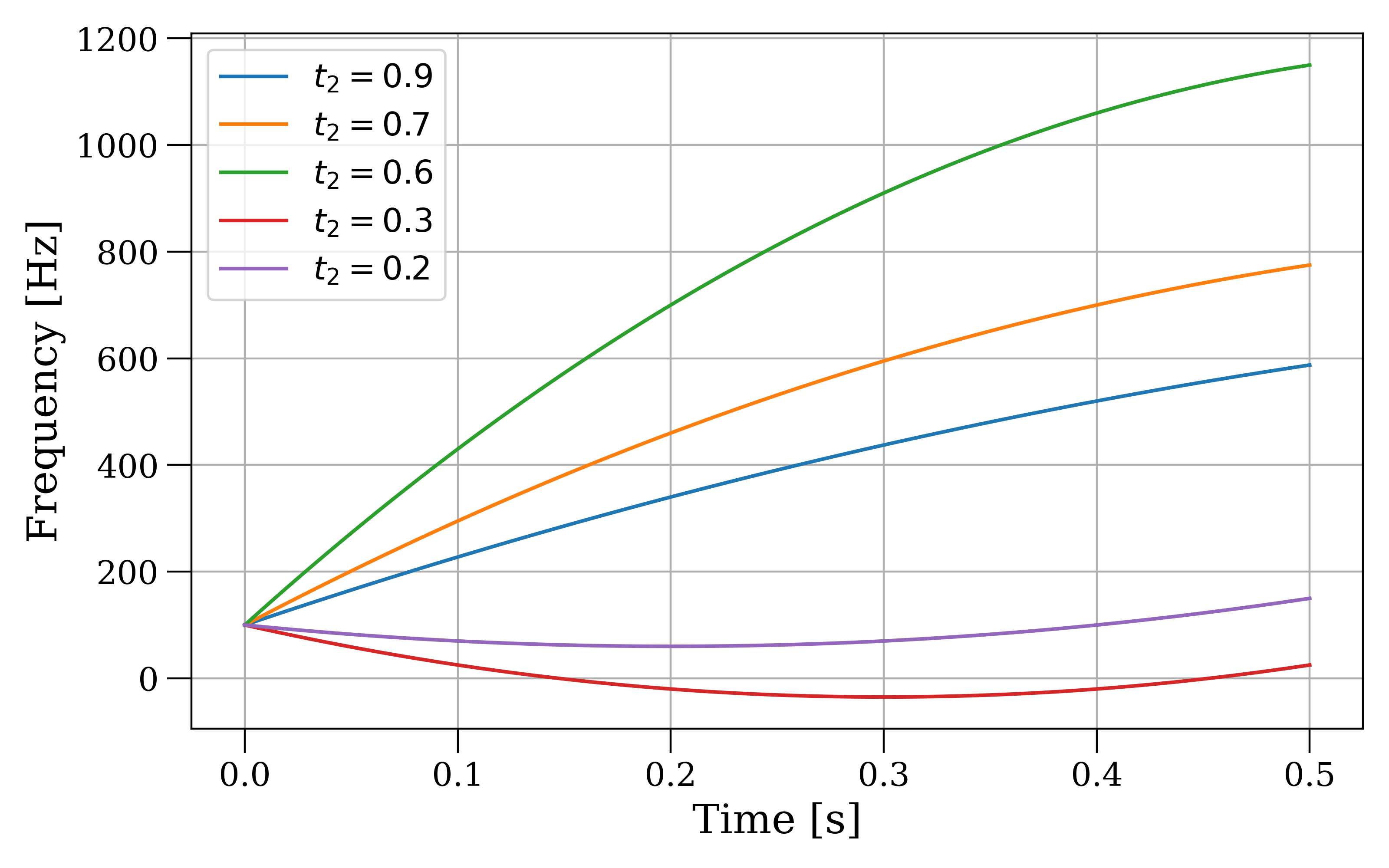}
      \caption{\footnotesize{
    Examples of the evolution of the frequency $f(t)$ for different values of $t_{2}$ with $t_{2}>t_{end}$  and $t_{2}< t_{end}$. The change in $t_{2}$ represents how this variable controls the concavity of our frequency function. When the value of $t_{2}$ is higher than $t_{end}$, the rising arch is obtained for the HFF (green, blue, and orange lines).  
      }}
    \label{fig:t_2 plots}
 \end{figure}

\begin{figure}
    \centering
    \includegraphics[scale=0.5]{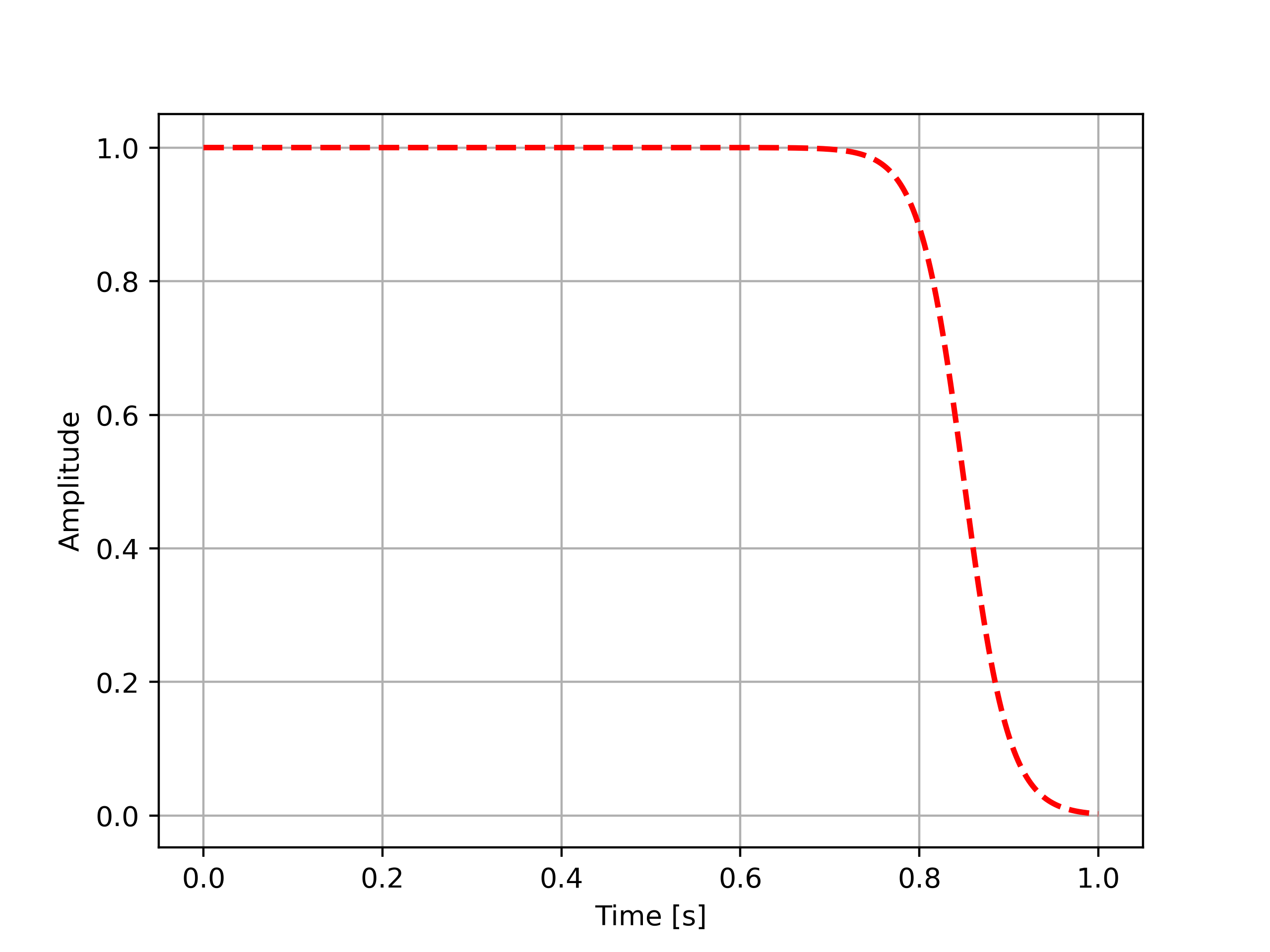}
    \caption{\footnotesize {Logistic Regression function with a middle point in $x_{0} = 0.8$ and a steepness of $k = 40$. Using equation \eqref{Eq: Logistic_misdirec} we plot the dot lines that show the modified logistic regression function that allows smoothing the final part of the signal.}}
    \label{fig:Sigmoid}
\end{figure}

%
The arch produced by the evolution of the HFF in CCSN GW sinals is modeled by the angular frequency $w(t) = 2\pi f(t)$, 
to determine an analytical model that fits these needs for the function $f(t)$ following \cite{astone2018new}, for this we introduce a second-degree polynomial model that satisfies:
\begin{equation}\label{E:Freq_mod}
    f(t) = f_{0} + f_{1}(t-t_{ini}) + f_{2}(t - t_{ini})^{2} \quad ; \quad t \in [t_{ini}, t_{end}],
\end{equation}
where $f_{0}$ is a constant value, $f_{1}\, \text{and} \, f_{2}$ are time functions which control the frequency evolution. The time interval will be from the initial time $t_{ini}$ to the final time $t_{end}$. The appropriate ramp-up function according to the simulations given by \cite{cerda2013gravitational}, is expressed by:
\begin{align*}
   f(t) & = f_{0} + \frac{2(f_{1s} - f_{0})(t_{2}-t_{ini})}{(2t_{2} - t_{ini} - 1)(1 - t_{ini})}(t - t_{ini}) \\
   & -  \frac{f_{1s} - f_{0}}{(2t_{2} - t_{ini} - 1)(1 - t_{ini})}(t - t_{ini})^{2},
\end{align*}
where $f_{1s}$ is defined as the frequency evolution at one second $f(t = 1)$, $t_{2}$ influences the concavity of the frequency curve and represents the maximal point of the polynomial where $t_{2} \geq t_{end}$. The values of $t_{2} \ll t_{end}$ alter the orientation of the curvature so that the maximum value of $f(t)$ increases faster than expected in the simulations.
Finally, the driven frequency is defined through the number of accelerations $N$, given by equation \eqref{E:Forcing} as:
\begin{equation} \label{E:f_driven}
    f_{driven} \equiv \frac{N}{t_{end} - t_{ini}},
\end{equation}
this frequency characterize the number of triggers per unit time introduced by the forcing. 
In Fig.~\ref{fig:t_2 plots} we plot the frequency function equation \eqref{E:Freq_mod} with a starting point of $0$ to $0.5$ s with a frequency from $-50$ to $1200$Hz.
The change in $t_{2}$ represents how this variable controls the concavity of the frequency function $f(t)$. When $t_{2}\gg t_{end}$ approaches the left, nearly to $t_{end}$, we will start to have negative values that are not allowed by the HFF. When the value of $t_{2}$ is greater than $t_{end}$, we will observe the rising arch estimated to obtain the HFF behavior.
The analysis of generated CCSN GW signals is carried out using the analytical model detailed in Section \ref{section: Generation of GW}. This model allows for the customization of various characteristics of the signal, such as the density of amplitude points, duration, and the linear increase of frequency pixels over time. However, the behavior of the signal in the last milliseconds of its duration, particularly its smoothness during the last phase of evolution, cannot be directly controlled. This lack of control introduces significant complexities in signal analysis. To address this issue, we propose the incorporation of the logistic function, a smooth, continuous, $S$ shaped curve \cite{cramer2002origins} that gradually decreases during the final moments of the generated signals. At this stage, the amplitude signal becomes zero, which means that the absence of oscillations generated in the PNS is responsible for the emission of GW.
This logistic function is represented by the following equation:
\begin{equation}
    S(x) = \frac{1}{1 + e^{-k(x-x_{0})}} = \frac{e^{k(x-x_{0})}}{1 + e^{k(x-x_{0})}},
    \label{Eq:Logistic_reg}
\end{equation}
where $x_{0}$ denotes the middle point of the $S$ curve and  $k$ indicates the steepness of the function.
Thus, to achieve a smooth transition of the signal at the final time of the waveform, it is essential to modify the function $S(x)$ in equation \eqref{Eq:Logistic_reg} by incorporating a negative sign to enable the desired inversion of the function. This subtraction will then be used to create a "misaligned" logistic function $T(x)$ in the following manner:
\begin{equation}
    T(x) = 1 - S(x).
    \label{Eq: Logistic_misdirec}
\end{equation}
The modified logistic regression function in equation \eqref{Eq: Logistic_misdirec} is used to multiply the generated waveform and allow smoothing of the final part of the signal; the function form can be seen in Fig.~\ref{fig:Sigmoid}, where the middle point is reached in $x_0=0.8$ using a steepness with a value equal to $k=$40. 
%
\subsection{The semi-implicit Euler method}
To derive a solution $h(t)$ from the equation \eqref{E:Driving} due to the lack of an analytical solution, we employ a basic numerical technique known as the semi-implicit Euler method.
This method relies on a finite difference approximation, enabling the transformation of the differential equation into a linear recurrence. To solve the equation, the order is initially reduced by redefining the variables in the following way:
\begin{equation}    
x_{1}(t) = h(t) \qquad {\rm and} \qquad x_{2}(t) = \frac{d h(t)}{dt},
\end{equation}
by differentiating $x_2$ we obtain a first order equation to solve,
\begin{equation}
    \frac{d x_{2}(t)}{dt} = s(t)  - \frac{w(t)}{Q}x_{2} - w(t)^{2}x_{1}(t). 
    \label{Eq: SmE_x}
\end{equation}
Then, using this finite differential method to solve numerically our equation, we describe the variables and derivatives in terms of recurrence relations,
\begin{equation}
    \begin{split} 
    x_{2}(i+1) &= x_{2}(i) + \left( s(i)  - \frac{w(i)}{Q}x_{2} - w(i)^{2}x_{1}(i) \right)\Delta t,\\
    x_{1}(i+1) &= x_{1}(i) + x_{2}(i+1) \Delta t,
    \end{split}
\end{equation}
where $i=0...a$ is a constant that determines the next step of the simulation. The following equations \eqref{E:algoritmo} describe the semi-implicit Euler method used in the simulation. 
\begin{align}\label{E:algoritmo}
dh[i + 1] &= dh[i] + (s[i] - (w[i]/Q) (dh[i]) - w[i]^{2}(h[i]))dt ) ,\nonumber  \\  
h[i + 1]  &= h[i] + (dh[i + 1])dt.
\end{align} 
The accurate determination of the parameters that govern the differential equation \eqref{E:Forcing}, significantly influences the modeling of the generated signals. The parameters implemented to generate different responses are shown in Table \ref{T:Parameters}. With these parameters, we obtain different linear growth patterns over time; equivalently, this defines different slopes of the HFF.
%
 \begin{table*}
	\centering
	\begin{tabular}{lcc}	
	\hline 
	\multicolumn{3}{c}{\emph{\text{Model Parameters}}}  \\ 
	\hline 
	$\mathbf{Parameters \qquad}$  & \textbf{Minimum } & \textbf{\qquad Maximum} \\ \hline 
        ${\bf t}_{ini}$[s]  & 0.0   & \qquad 0.35   \\ 
        ${\bf t}_{end}$[s]  & 0.3   & \qquad 1.0    \\ 
        ${\bf f_{0}}$[Hz]   & 100   & \qquad 650    \\
        ${\bf f_{1s}}$[Hz]  & 450   & \qquad 2000   \\
        ${\bf t_{2}}$[s]    & $0.35$& $\qquad  1.5$\\
        ${\bf Q}$    & $0.5$  & \qquad $100$  \\
        ${\bf f_{driver}}$[Hz]& 200 & \qquad 400    \\
        ${\bf f_{s}}$ [KHz] & 4  & \qquad 16\\ \hline
	\end{tabular}
	\caption{\footnotesize Parameters that rule our generated waveforms and compose the parameterized frequency function. The table shows in the first column, the parameter space of the generated templates, the second and third columns indicate the range (minimum and maximum, respectively) for each parameter.}
 \label{T:Parameters}
\end{table*}

Besides the basic parameters, it is essential to establish the following: the quantity of triggers $N$ using the formula~\eqref{E:f_driven}, the highest magnitude $s_{n}$ as indicated in equation \eqref{E:Forcing}, and the parameters for the logistic function, specifically, the midpoint $x_{0} = 0.8$ and the slope $k = 40$ from equation~\eqref{Eq: Logistic_misdirec}. Once these values are determined, we can proceed with the computational simulations of the waveforms, which will be applied in the subsequent section.
%
\section{Results \label{section: Results}}
%
In this section, we provide the outcomes of applying this approach to reconstruct simulated CCSN GW signals. In addition, we determined the effectiveness of the reconstructed signals compared to the LIGO noise. 
%
\subsection{Algorithm Implementation: CCSN GW spectrogram production} 
\label{subsection: Algthm imp}
\begin{figure*}
    \centering
    \includegraphics[scale=0.58]{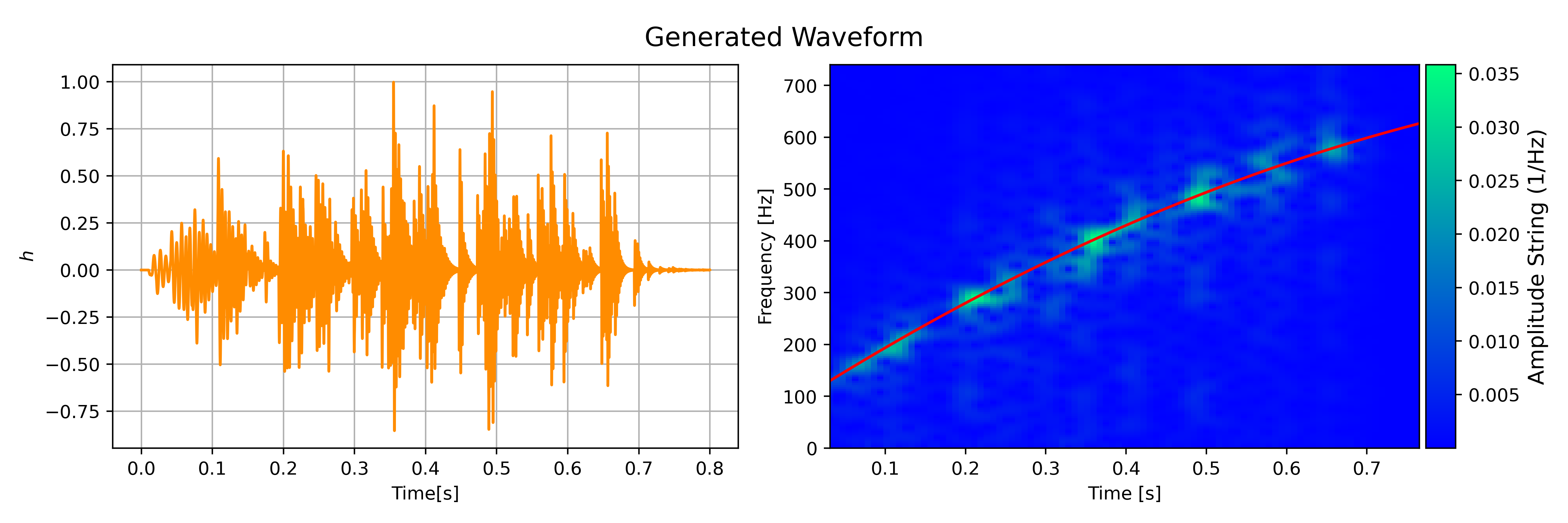}
    \caption{{
    The left panel displays the strain signal of GW CCSN, it is highlighting stochastic nature throughout its duration. The right panel shows the spectrogram with the HFF evolution, demonstrating a linear and continuous increase in frequency, aligning with the expected behavior. The signal has a duration of $0.8$ s with a maximum frequency of $600$ Hz.
    }}
    \label{fig:single}
\end{figure*}

\begin{figure*}
    \centering
    \includegraphics[scale=0.6]{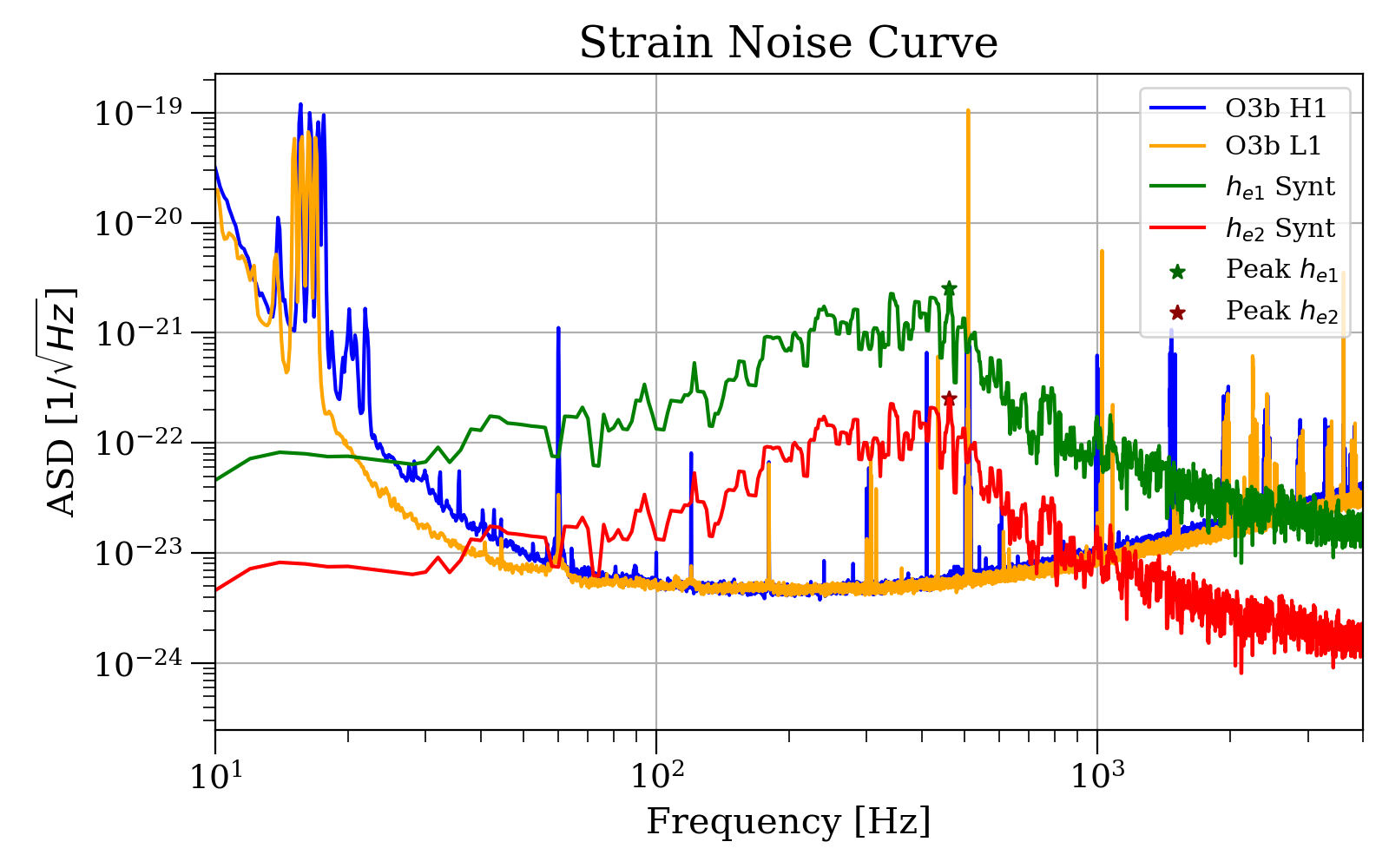}
    \caption{Noise amplitude curve obtained during the scientific run O3b of LIGO interferometers L1 and H1. 
    The plot shows how two different generated CCSN GW signals ($h_{e1}, h_{e2}$) respond to the sensitivity of laser interferometric detectors.}
    \label{fig:ASD1}
\end{figure*}

%
To derive the frequency function $f(t)$ as described in equation \eqref{E:f_driven} and to replicate the linear growth in the HFF, the coefficient values specified in Table \ref{T:Parameters} were used. Subsequently, through a comprehensive implementation of the different steps introduced in the algorithm outlined in equation \eqref{E:algoritmo}, the waveform $h(t)$ was generated, ensuring its compatibility with the LIGO interferometer detection range. This waveform was then normalized to accommodate any desired amplitude and resampled at the sample rate of the LIGO detector ($f_{s} = 16384$) \cite{ligo2020guide}, as shown on the left side of Fig.~\ref{fig:single}.
The signal spectrogram is shown on the right side of Fig.~\ref{fig:single}, we can observe the linear and continuous increase in frequency, aligned with the expected HFF behavior of the signal discussed. The signal has a duration of $0.8$s with a maximum frequency of $600$Hz. 
To align the modeled signals with the sensitivity range of L1 and H1 interferometers, the Amplitude Spectral Density (ASD) is computed for a specific waveform model \cite{moore2014gravitational}. By applying a scaling factor to the normalized amplitude of the waveform, different ASD profiles are derived, each corresponding to a different range of sensitivity. Two of these ASD profiles for the generated signals ($h_{e1}, h_{e2}$) are compared with the ASD data from the LIGO observation run (O3b) illustrated in Fig.~\ref{fig:ASD1}, demonstrating how this scalable model can operate effectively within the detection thresholds of these interferometers.


\subsection{Generation of CCSN GW signals} \label{section: GW CCSN}
\begin{figure*}
    \centering
    \includegraphics[scale=0.57]{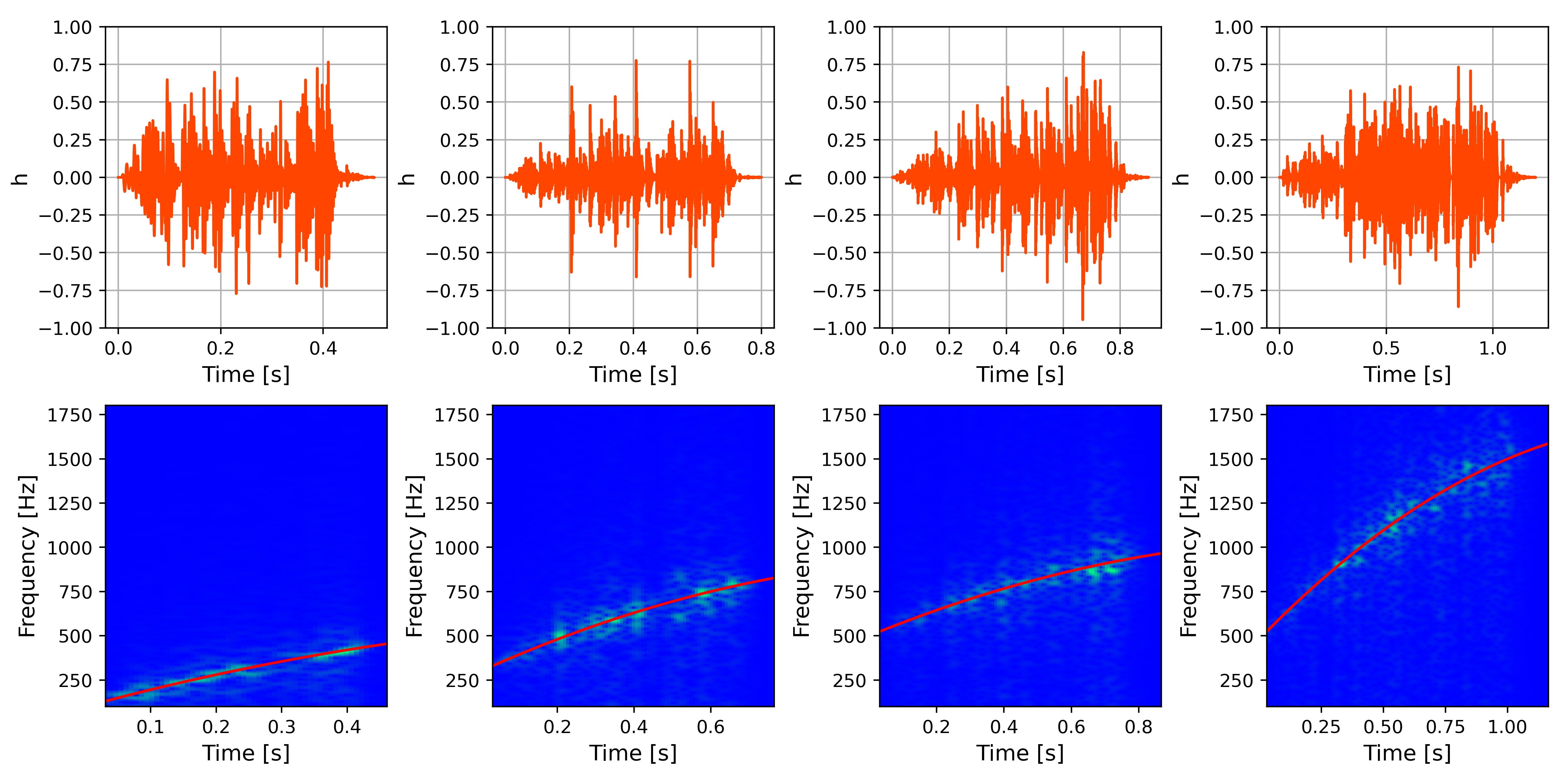}
    \caption{The upper panel displays the strain data for WF1, WF2, WF3, and WF4 as detailed in Table \ref{T:Parameters2}. These graphs demonstrate a significant random element, influenced by the stochastic forcing elucidated in the analytical model presented in Section \ref{subsection: Phen gmode -model}. The lower panel illustrates the time-frequency evolution plot for each signal generated. Within each spectrogram, the rising slope of the high-frequency content is observable, showcasing how this slope fluctuates due to adjustments made to the parameters governing the high-frequency content evolution.}
    \label{fig:WF_4signal}
\end{figure*}

%
We have generated 1000 waveforms with varying HFF ranges between $1600,3500$~Hz. These waveforms differ in strain duration, initial frequency, and the driven frequency, which are restricted by the limits specified in Table \ref{T:Parameters}. Table \ref{T:Parameters2} displays four waveform samples (WF1, WF2, WF3, WF4) with different representative parameters.
 \begin{table*}
	\centering
	\begin{tabular}{ccccccccc}	
	\hline 
	\multicolumn{9}{c}{\emph{\text{Initial Values}}}  \\ 
	\hline 
	WF & $\mathbf{t_{ini}[s]}$ & $\mathbf{t_{end}[s]}$ & $\mathbf{f_{0}[Hz]}$&$\mathbf{f_{1s}[Hz]}$ & $\mathbf{t_{2}[s]}$ & $\mathbf{Q-factor}$ & $\mathbf{f_{driver}[Hz]}$ & \textbf{HFF Slope}\\ 
    \hline
	$\mathbf{WF1}$	&  0.0  &  0.5 &  100 &  600 &  1.5 &  10 &  400 & 750.17  \\
	$\mathbf{WF2}$	&  0.0  &  0.8 &  300 &  900 &  1.5 &  10 &  400 & 675.05  \\
	$\mathbf{WF3}$	&  0.0  &  0.9 &  500 &  1000 &  1.5 &  10 &  400 & 527.83  \\
	$\mathbf{WF4}$	&  0.0  &  1.2 &  480 &  1500 &  1.5 &  10 &  400 & 935.08  \\\hline
	\end{tabular}
	\caption{\footnotesize Parameters associated to generate CCSN GW signals and how different configurations of these parameters combines to produce differnt linear responses in the HFF. We compute such responses in the spectrogram of the signal estimating the slope of the HFF using the simple relation HFF$_{slope}=f_{1s}-f_0/t_{2}-f_{ini}$.}
    \label{T:Parameters2}
\end{table*}
The fluctuations in the HFF slope's reaction demonstrate the model's capability to produce CCSN GW signals similar to those shown in Fig.~\ref{fig:WF_4signal}, and how the slope can vary across all spectrograms.
%
\section{LIGO response associated to CCSN GW signals from generated and numerical simulations\label{subsection: 3D CCSN}}
\begin{figure*}
    \centering
    \includegraphics[scale=0.40]{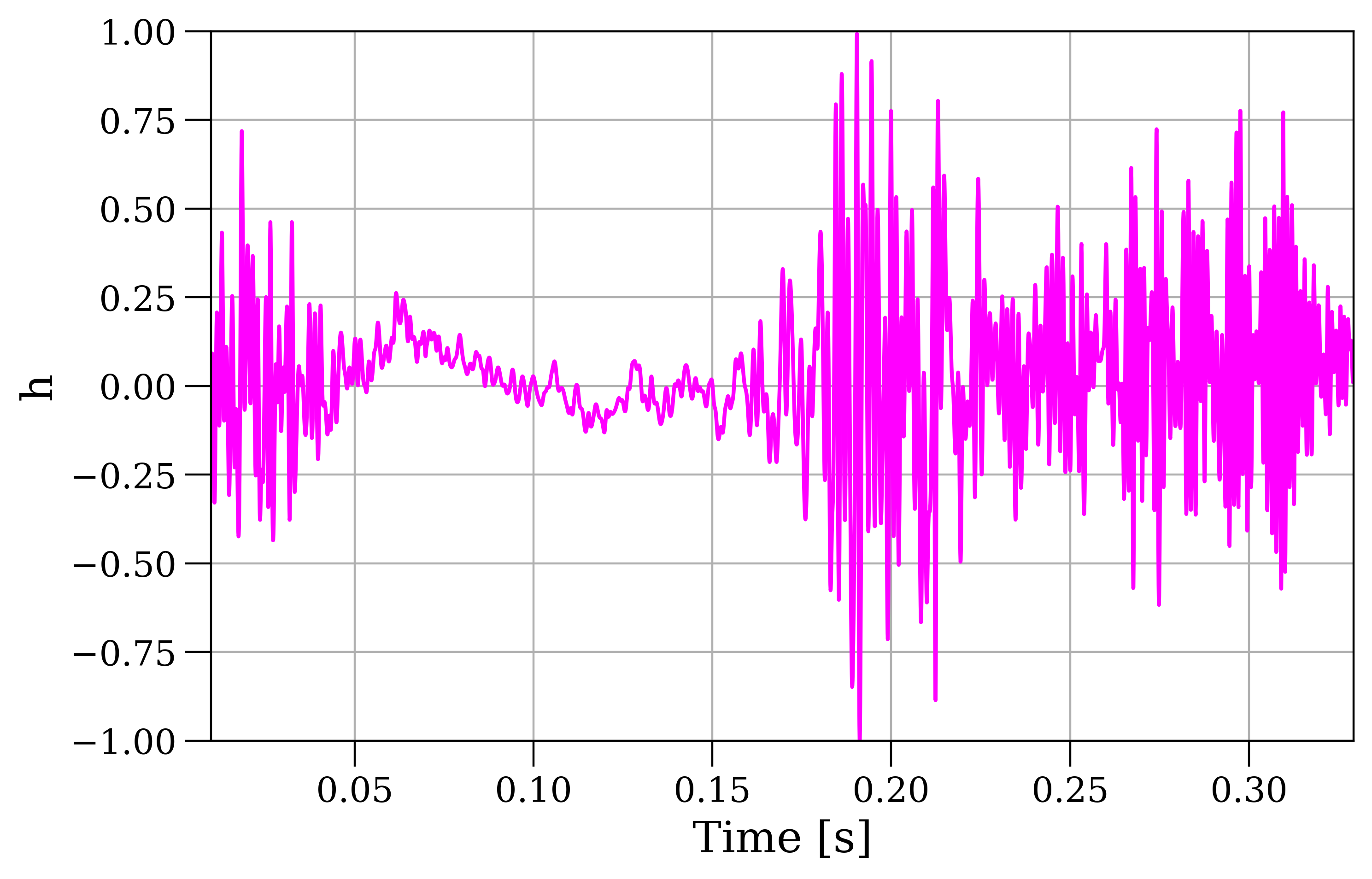}
    \includegraphics[scale=0.40]{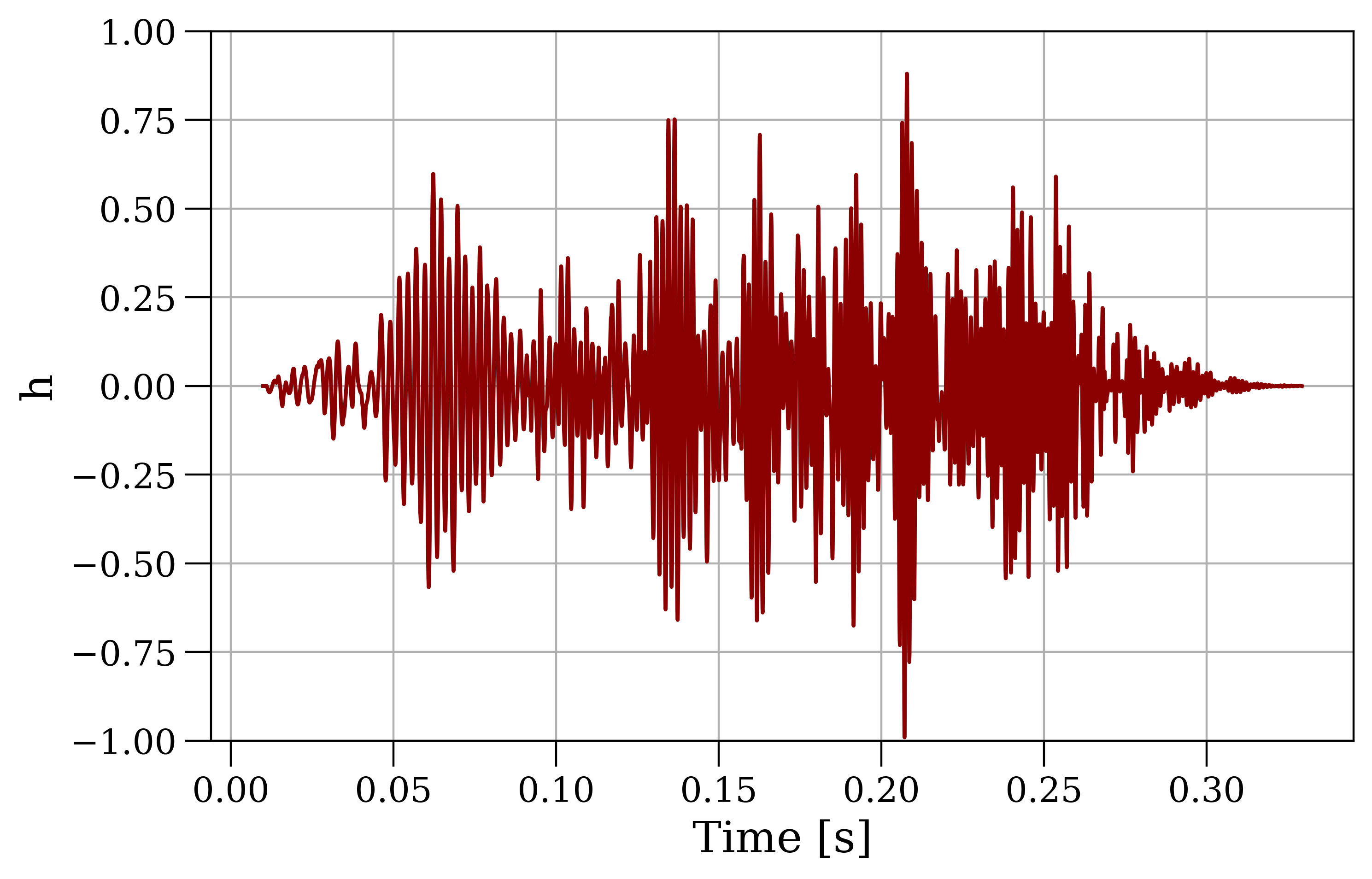}
    \caption{Left panel Strain signal for Andresen et al. (2017), model s15.nr. Right panel Strain of a generated CCSN GW signal.}
    \label{fig:ASD}
\end{figure*}

\begin{figure*}
    \centering
    \includegraphics[scale=0.40]{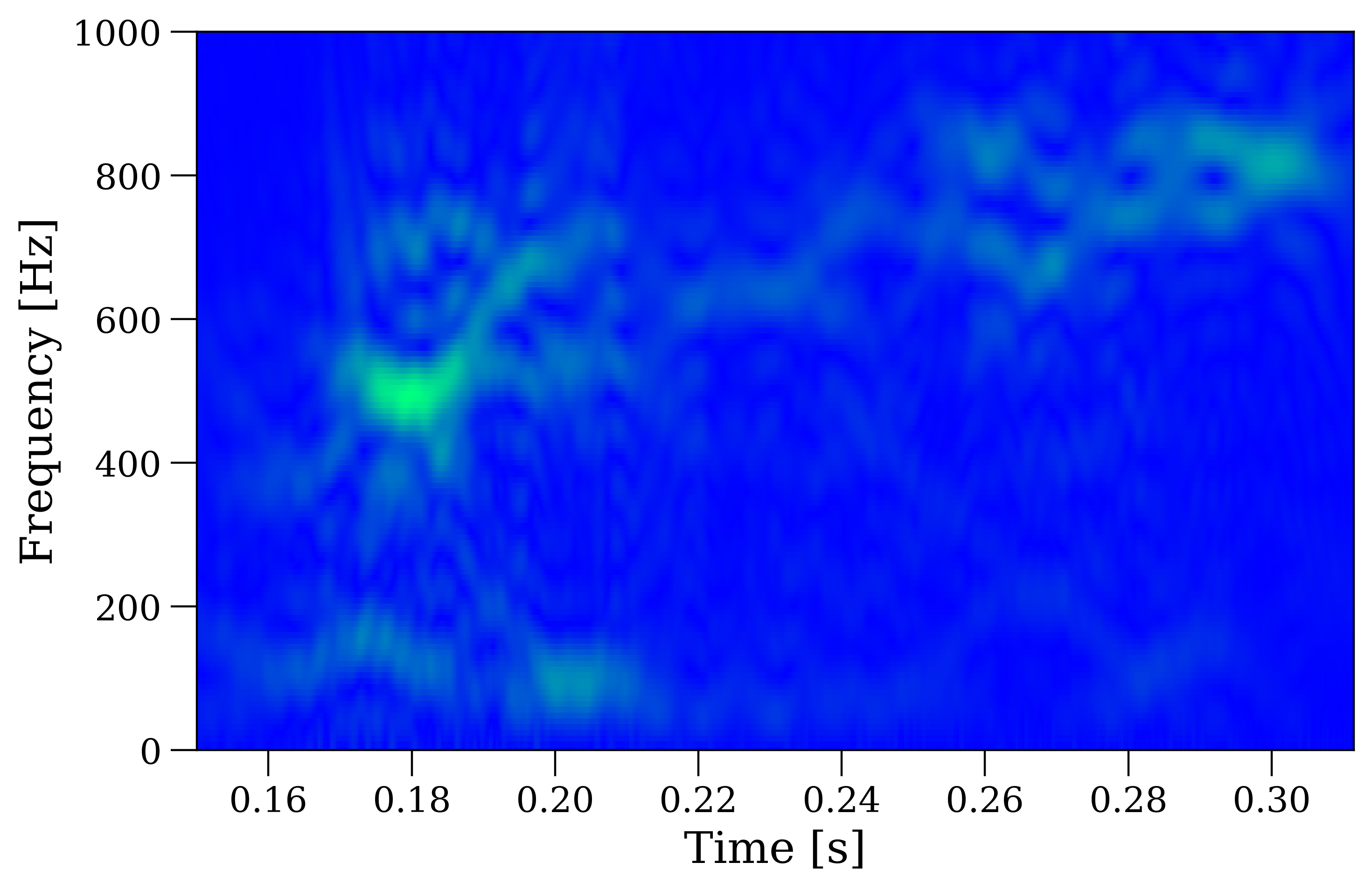}
    \includegraphics[scale=0.40]{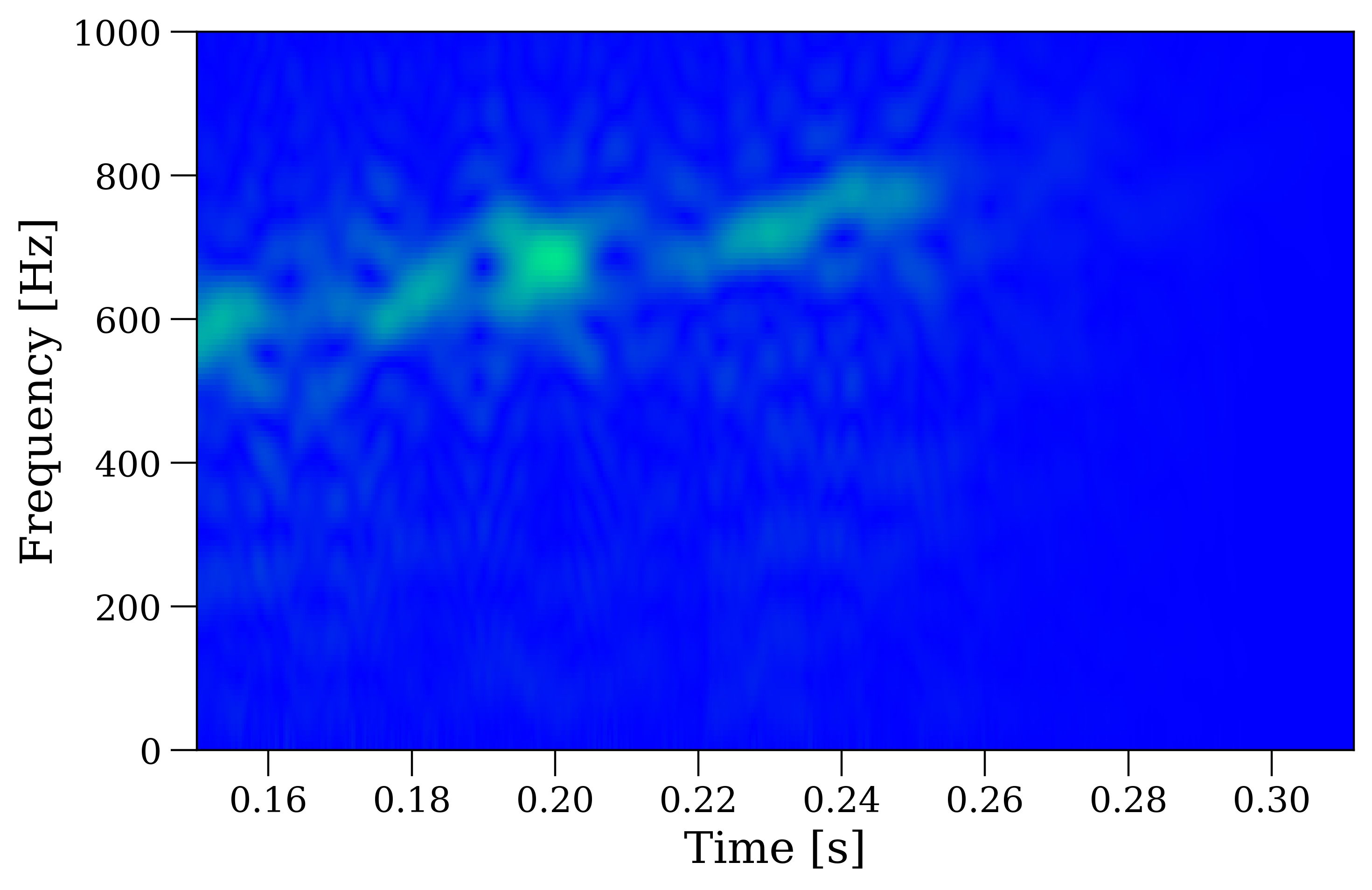}
    \caption{Spectrograms of CCSN GW signals. Left panel show the time-frequency evolution plot for Andresen, model s15.nr. Right panel
    GW signal associated with a generated CCSN GW signal obtained using the methodology described in Section \ref{section: Generation of GW}. 
    }
    \label{fig:spectrogram2}
\end{figure*}

\begin{figure*}
    \centering
    \includegraphics[width=0.65\linewidth]{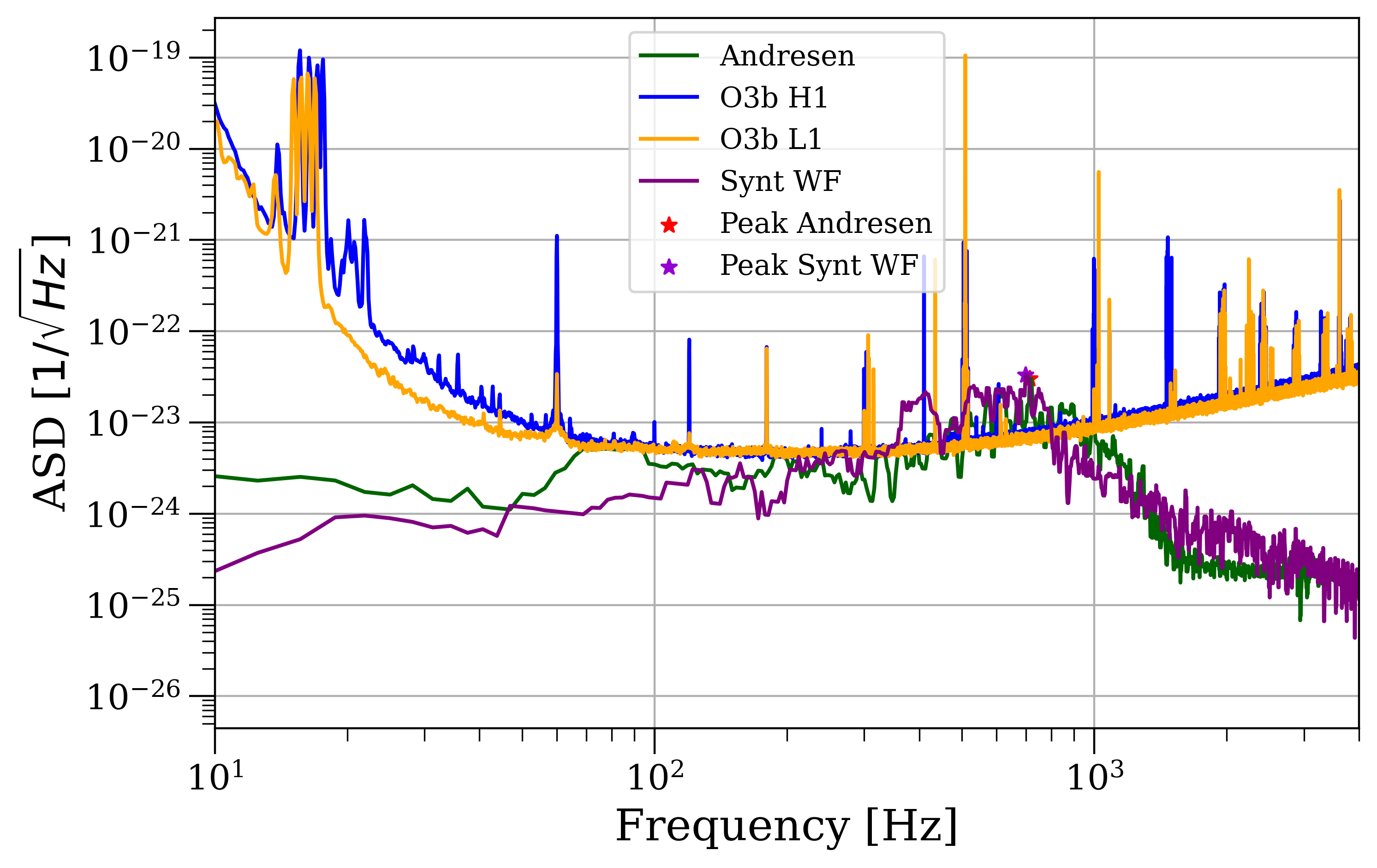}
    \caption{
    In the figure, we compare the ASD of the Andresen signal (green), and a generatedwaveform (purple) with respect to the L1 and H1 LIGO interferometers highlighted in yellow and blue respectively. The maximal peak for Andersen and Synt WF are indicated with a star.}
    \label{subfig:CCSNe}
\end{figure*}

%
To investigate the physical features of our generated CCSN GW signals, we compare the strain signals, time-frequency evolution spectrograms, and GW-sensitive curves related with a 3D CCSN GW numerical signals. We aim to evaluate the appropriateness and consistency of different physical characteristics in signal generation for use in LIGO interferometric research, including differences in duration, peak frequency, and amplitude spectral density.
To make such a comparison, we use the 3D CCSN GW signals from Andresen et al. 2019 \cite{Andresen_2017} model \textit{ s15.nr}. These GW signals are extracted from three different models based on approximately 3D general relativistic radiation (neutrino) hydrodynamic simulations with a single progenitor with a zero-age main-sequence mass (ZAMS) of 15~$M_{\odot}$, solar metallicity, and with different rotation rates $0$\;rad/s, $0.2$\;rad/s, and $0.5$\;rad/s. 
We select this model because it represents the prototype of 3D CCSN numerical simulation, which contains the most studied progenitor mass, with physical properties such as angular momentum and HFF presence, which cover some of the basic elements in the numerical simulation analysis.  
In continuation of the previous approach, a randomly generated waveform is chosen to analyze the strain of the generated CCSN GW signal against the model \textit{s15.nr}. The comparison was carried out considering the following features:
\begin{enumerate}
    \item \textbf{\textit{GW Strain:}} The GW strain signal from Andresen model \textit{s15.nr} is characterized by a stochastic behavior, similar in form to those obtained for the generated model as is illustrated in Fig.~\ref{fig:ASD}. These strain signals and the generated CCSN GW share the characteristic of stochastic behavior in this type of emission.  
    \item \textbf{\textit{GW Spectrogram:}} The time-frequency evolution for these two signals exhibits a random distribution of frequencies over time, the characteristic arch that defines the slope evolution can be identified. In the generated model, this feature is consistent with the properties described in Section \ref{section: Generation of GW}, producing a linear increase in time see Fig. \ref{fig:spectrogram2}.  
    \item \textbf{\textit{GW sensitivity:}} The sensitivity curves presented in Fig.~\ref{subfig:CCSNe} reflect similar noise amplitudes for generated and 3D CCSN GW signals. Some differences can be found near $10^{2}$~Hz; however, the response of the signals is quite similar. 
\end{enumerate}
As we can see, the CCSN GW signals replicate the stochastic component during its temporal evolution, as required for these signals. 
Finally, the comparison of ASD between the Andersen model s15.nr (shown in green) and the corresponding generated CCSN GW signal (depicted in purple) is illustrated in Fig. \ref{subfig:CCSNe}. This comparison serves to demonstrate that, by adjusting the parameters that govern the theoretical model, we are able to scale the waveform response within the sensitivity range of LIGO interferometers, thus shifting the generated CCSN GW signals as depicted in Fig. \ref{fig:ASD}. It can be inferred that both the Andersen and the generated CCSN GW signals fall within the detection ranges of the H1 and L1 interferometers and can be fine-tuned to meet specific detection criteria.
%
\section{Conclusions}
\label{Conclusions}
%
In this study, a computational model was created based on the physical characteristics of a damped harmonic oscillator driven by force $s(t)$, to generate CCSN GW signals more efficiently than traditional numerical simulations. We show that the temporal development of the signals produced shows a linear increase in frequency. It can be modified by adjusting the parameters that govern the dynamics of the physical system by emulating the high-frequency features found in all simulated CCSN GW signals. This provides an opportunity to estimate the slope of the high-frequency features more accurately, as the linear growth simplifies the calculation of this parameter compared to the model proposed by Andresen et al. in 2018.
The application of the theoretical framework suggests that we can replicate the random patterns of the CCSN GW signals by incorporating a driving force, represented by a Dirac delta function. Of particular significance, the graphs in Figure \ref{fig:ASD} show a strong similarity in terms of detectability and sensitivity in the strain noise curves. This similarity confirms that the produced signals are suitable for use in research involving the estimation of high-frequency features in actual interferometric noise.
\\ \\
As anticipated, this approach paves the way for the estimation of parameters in CCSN GW signals by incorporating subsequent measurement of the HFF slope, which could be beneficial for future research efforts, such as refining parameter estimation techniques and improving data analysis methods \cite{PhysRevD.108.084027}. 
\\ \\
The subsequent steps for parameter estimation when a CCSN signal is detected in real noise involve implementing the generated signals for use in cWB event production analyses to create, for example, a training data set with estimated HFF slope values in a machine learning framework suitable for LIGO interferometric noise (see Casallas-Lagos et al. 2023). This research can be expanded to highlight the influence of specific physical parameters of the source in the estimation of the slope of the HFF, such as the nuclear equation of state. This extension will be addressed in a subsequent paper, scheduled for publication in spring 2024, continuing the work of Casallas-Lagos et al. (2023).
By comparing the signals simulated in this research with the detector sensitivity curve during the LIGO run O3b (refer to Figure \ref{fig:ASD}), it can be deduced that the generated signals reach the same detection threshold as the numerical model, thus validating the usefulness of using the generated CCSN
GW signals in investigations that explicitly consider interferometric noise.
\\ \\
The results achieved through the application of the methodology suggested in this study demonstrate the feasibility of identifying particular deterministic elements within a CCSN GW signal, such as the HFF. The signals generated by CCSN GW possess the capability to serve as a versatile and cost-effective tool that could complement LVK interferometric data for extracting the physical information embedded in the deterministic characteristics included in the stochastic CCSN GW emission. As a next step, it would be intriguing to replicate waveforms that encompass a broader range of physical parameters found in a GW produced by CCSN.

\section{Acknowledgements}
This work was supported by CONAHCYT Network Project No. 376127 Sombras, lentes y ondas gravitatorias generadas por objetos compactos astrofísicos. A.C.L. and C.T. acknowledge the CONAHCYT scholarship. C.M. thanks CONAHCYT and PROSNI-UDG.

\bibliographystyle{unsrtnat}
\bibliography{references}  

\end{document}